\documentclass[aps,preprintnumbers, notitlepage,twocolumn]{revtex4}

\usepackage{graphicx}%
\usepackage{dcolumn}
\usepackage{bm}

\usepackage{subfig}
\usepackage{color}
\usepackage{amsmath}
\usepackage{amsfonts}
\usepackage{slashed}
\newcommand{\Feyn}[1]{#1\kern-0.45em/}

\begin{document}
\preprint{LA-UR-13-22756}
\title{MeV dark matter in the $3+1+1$ model} 

\author{Jinrui Huang}
\email{jinruih@lanl.gov}
\affiliation{Theoretical Division, T-2, MS B285, Los Alamos 
National Laboratory, Los Alamos, NM 87545, USA}
\author{Ann E Nelson}
\email{aenelson@u.washington.edu}
\affiliation{Department of Physics, University of Washington, Seattle, Washington 98195-1560, USA}

\date{\today}
\begin{abstract}
\label{abstract}
The existence of light sterile neutrinos in the eV mass range with relatively large mixing  angles with the active neutrinos has been proposed for a variety of reasons, including to improve the fit to the LSND and MiniBooNE neutrino oscillation experiments, 
and reactor disappearance experiments. In ref.~\cite{Nelson:2010hz}, it was shown that neutrino mixing with a heavier sterile neutrino, in the mass range between 33 eV and several GeV, could significantly affect and improve the agreement between neutrino oscillation models with light sterile neutrinos and short baseline experimental results, allowing for a new source of CP violation in appearance experiments and for different apparent mixing angles in appearance and disappearance experiments. However in refs.~\cite{Kuflik:2012sw} and \cite{Fan:2012ca} it was shown that various collider experiment, supernovae, and cosmological constraints can eliminate most of the parameter region where such a heavy sterile neutrino can have a significant effect on neutrino oscillations. In this paper we consider the effects of allowing a new light scalar in the MeV mass region, which is a potential dark matter candidate, to interact with the sterile neutrinos, and show that the resulting model is a consistent theory of   neutrino oscillation anomalies and dark matter which can also potentially explain the INTEGRAL excess of 511 keV gamma rays in the central region of the galaxy.
\end{abstract}

\maketitle

\section{Introduction}
\label{sec:intro}
Sterile neutrinos--neutrinos which do not have standard model electroweak interactions, could mix with the active neutrinos and affect neutrino oscillation experiments. Several anomalies, such as  
the excesses of electron antineutrinos observed at  the LSND~\cite{Aguilar:2001ty}  and  MiniBooNe~\cite{AguilarArevalo:2007it, AguilarArevalo:2008rc, AguilarArevalo:2010wv} short baseline neutrino oscillation experiments, the excess of electron neutrino candidate events observed at MiniBooNE, the Antineutrino Reactor Anomaly~\cite{Mention:2011rk, Huber:2011wv} have been interpreted as evidence for sterile neutrinos. An additional light state also can improve the fit to CMB and nucleosynthesis cosmology~\cite{Dunkley:2010ge, Hou:2012xq, Izotov:2010ca, Aver:2010wq, Hamann:2010bk,DiValentino:2013qma,Said:2013hta}. A minimal model with one sterile neutrino, ``$\nu_4$"  in the eV mass range, the so-called $3+1$ model, does not give a good fit to all the short baseline neutrino oscillation data ~\cite{Kopp:2011qd, Maltoni:2007zf}. More detailed studies and various cosmological and collider constraints have also been investigated in the paper~\cite{Kuflik:2012sw} and it turns out no allowed parameter space has been found for  mixing angles between the sterile neutrinos and active neutrinos large enough to fit the LSND and MiniBooNe results. The  addition to the $3+1$ model of a sterile neutrino, ``$\nu_5$", which is heavier than 33 eV, the ``$3+1+1$ model"~\cite{Nelson:2010hz}, is interesting in that even though oscillations involving such a heavy neutrino have an unobservably short wavelength, after averaging over the short wavelength oscillations the effective theory still allows for significant CP violation in appearance experiments and furthermore allows for the effective mixing angles governing the oscillation amplitudes to be different in  neutrino oscillation appearance and disappearance experiments. However, combing the data from Planck~\cite{Ade:2013ktc}, nine year WMAP~\cite{Bennett:2012zja,Hinshaw:2012aka} measurements, Baryon acoustic oscillations (BAO)~\cite{Percival:2009xn, Padmanabhan:2012hf, Blake:2011en, Anderson:2012sa, Beutler:2011hx} and the high-resolution ground-base CMB experiments: Atacama Cosmology Telescope (ACT)~\cite{Das:2013zf} and South Pole Telescope (SPT)~\cite{Reichardt:2011yv}, the recent PLANCK results~\cite{Ade:2013zuv} report the effective number of relativistic degrees of freedom $N_{eff} = 3.30_{-0.51}^{+0.54}$ at the $95\%$ confidence level (CL), which is consistent with three SM neutrino species, and an upper limit of 0.23 eV for the summed active neutrino masses. While including additional data from the direct measurements of the Hubble Constant $\text{H}_0$~\cite{Riess:2011yx}, the effective number of neutrino favors a higher value as $3.52_{-0.45}^{+0.48}$ at the $95\%$ CL. Nevertheless, there is still room for one additional sub-eV light sterile neutrino  with relatively large mixing angle with active neutrinos  ($m \sim 0.6$ eV), allowed by PLANCK~\cite{Ade:2013zuv}. In addition, a potential problem with a heavy sterile neutrino was pointed out in references \cite{Kuflik:2012sw} and \cite{Fan:2012ca} where it was argued that supernovae, nucleosynthesis, muon decay, and collider experiments constrain $\nu_5$ to have a mixing angle with the active neutrinos which is too small to significantly impact short baseline neutrino oscillation phenomena \cite{Aguilar:2001ty, AguilarArevalo:2007it, AguilarArevalo:2008rc, AguilarArevalo:2010wv}. 

In this paper we examine the effects of allowing the sterile neutrinos to have significant interactions with a light  ($\sim$ MeV) scalar. Light particles in the few MeV range have been proposed as part of a dark matter sector \cite{Boehm:2003hm,Pospelov:2007xh, Finkbeiner:2007kk, Cline:2010kv} and to explain the INTEGRAL (INTErnational Gamma-Ray Astrophysics Laboratory) excess of 511 keV gamma rays in the central region of the galaxy~\cite{Jean:2003ci} . Specifically, we introduce a Yukawa interaction between the two sterile neutrinos ($\sim \lambda \nu_4 \nu_5 \phi$, $\lambda$ is the coupling and $\phi$ is the singlet scalar), and assume that $\nu_5$ is heavier than the lighter sterile neutrino and $\phi$ field and can decay  into the other invisible states ($\nu_5 \rightarrow \nu_4 \phi$). The mass region of. $m_5 > m_{\pi} - m_{\mu} \sim 34$~MeV and $m_5 \lesssim 1$~MeV has been ruled out by various experiments shown in~\cite{Kuflik:2012sw}. Within the mass region of $m_{5} \sim [1,34]$~MeV, the short lifetime for $\nu_5$ can allow consistency with nucleosynthesis constraints. Although $\nu_4$, $\nu_5$ and $\phi$ all copiously produced in supernovae, if the interaction between them is large enough then all three of these new particles have short mean free paths in a supernovae and so the supernovae energy loss is dominated by standard model active neutrino emission.

The $\phi$ particle lifetime is very long, typically longer than the age of the universe, so it makes an interesting candidate for dark matter. With a mass in the MeV region, $\phi$ annihilation or decay into $e^+  e^-$ pairs can also explain the 511 keV gamma line observed by INTEGRAL~\cite{Boehm:2003bt}. Dark matter annihilation into $e^+e^-$ pair via the heavy charged particle exchange or via a new gauge boson was studied in ref.~\cite{Boehm:2002yz, Boehm:2003hm, Bai:2012yq}. For the simple case of light dark matter annihilating into the electron pair, to fit the 511 keV INTEGRAL gamma line as well as the continuum photon energy spectrum~\cite{Prantzos:2010wi}, the dark matter mass needs to be within the $\sim$[1, 30] MeV range.    Additionally, the dark matter mass needs to be less than 20 MeV, otherwise, the internal bremsstrahlung with positron production can violate the COMPTEL and EGRET diffuse gamma-ray observation~\cite{Beacom:2004pe}. More stringently, because it can also affect the fine structure constant, there is an upper bound 7 MeV on the mass of the dark matter particle~\cite{Boehm:2004gt}. Besides, the positron injection energy must be less than 3 MeV so that the gamma-ray spectrum from the positron inflight annihilation can be consistent with the diffuse Galactic gamma-ray data~\cite{Beacom:2005qv}. In the $3+1+1$ model, the $\phi$ particle mostly annihilates into sterile neutrinos and without introducing additional new interactions between the $\phi$ field and electron, while $m_{\phi} > 2m_{e}$, $\phi$ can decay into the electron and position pairs at the loop level and further explain the 511 keV INTEGRAL gamma line~\cite{Cline:2010kv}. Therefore, we are interested in the dark matter mass region of [2$m_e$, $m_5-m_4$] $\sim$ few MeV. 

The letter is organized as following. In Sec.~\ref{sec:one}, we describe the $3+1+1$ model in this letter and in Sec.~\ref{sec:one}, we search for the allowed parameter region which can avoid constraints from BBN and the supernovae SN1987A observations. We discuss the dark matter $\phi$ fields annihilating into sterile neutrinos and decaying into electron pairs and further explain the INTEGRAL 511 keV gamma-ray line in the Sec.~\ref{sec:three}. How to generate the realistic mass hierarchy and mixing for all of the SM fermions and sterile neutrinos through the $U(1)^{\prime}$ family symmetry is illustrated in the Sec.~\ref{sec:four}. We conclude in the Sec.~\ref{sec:five}.

\section{Models}
\label{sec:one}
\begin{figure}[t!]
\centering
\includegraphics[scale=0.4, angle = 0]{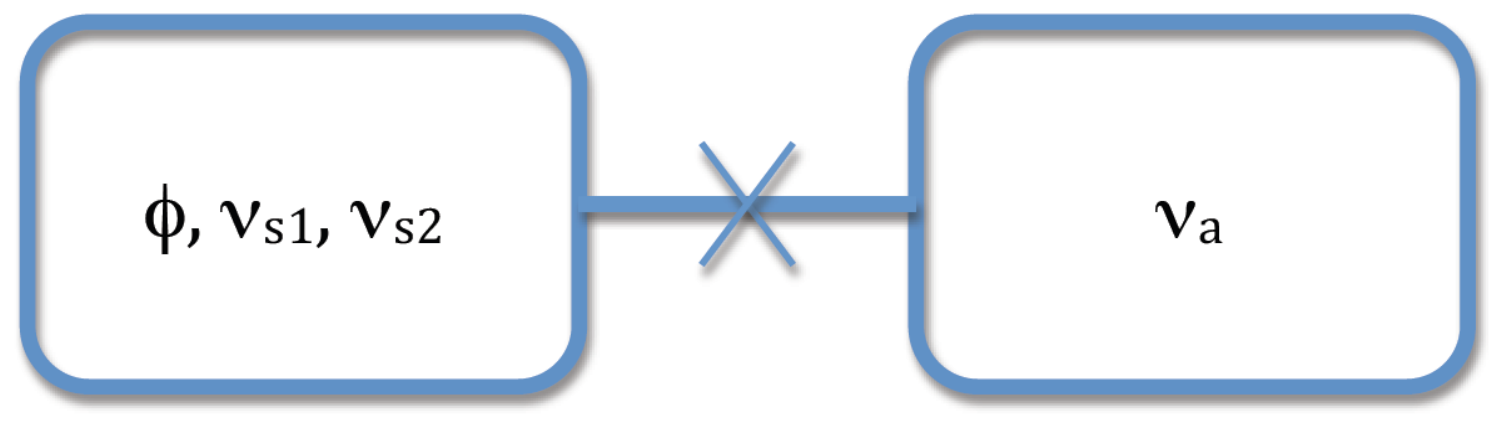}
	\caption{The cartoon picture illustrates the components of the dark sector and the interaction between the dark sector and visible sector. The box on the left is the dark sector while the box on the right represents the visible sector. The dark sector and visible sector interacts through the mixings shown as the cross symbol.}
	\label{fig:modelcartoon}
\end{figure}

Typically, sterile neutrinos are assumed to interact only via their mixing with active neutrinos. Here we consider the $3+1+1$ framework with an additional light scalar coupled to the sterile neutrinos. The Lagrangian corresponding to the new interaction is,
\begin{equation}
\mathcal{L}_{\phi} = \lambda \phi \nu_{s1} \bar{\nu}_{s2} + h.c. \;,
\label{eqn:lagrangian}
\end{equation}
where $\nu_{s1},\; \nu_{s2}$ are the additional sterile neutrinos in the gauge basis and dominant constituents of the $\nu_4$ and $\nu_5$ in the mass basis. We will use the approximation $\sim \lambda \phi \nu_4 \nu_5$ in the mass basis in the following sections. Effectively, $\phi\;, \nu_{s1}\;, \nu_{s2}$ stay in the dark sector and interact with our visible sector through neutrino mixing. It is depicted in the cartoon picture Fig.~\ref{fig:modelcartoon}.
In the dark sector, $\nu_5$ can decay invisibly through $\nu_5 \rightarrow \nu_4 \phi$ with the decay width of,
\begin{equation}
\Gamma_{\nu_5}^{\mbox{\tiny{inv}}} \simeq \frac{1}{16\pi} \lambda^2 m_5 \; ,
\end{equation}
and the light particles $\nu_4\;, \nu_5\;$ and $\phi$ also scatter among themselves.

\section{Experimental constraints}
\label{sec:two}
\subsection{BBN bound}
\begin{figure}
\centering
\includegraphics[scale=0.3, angle = 0]{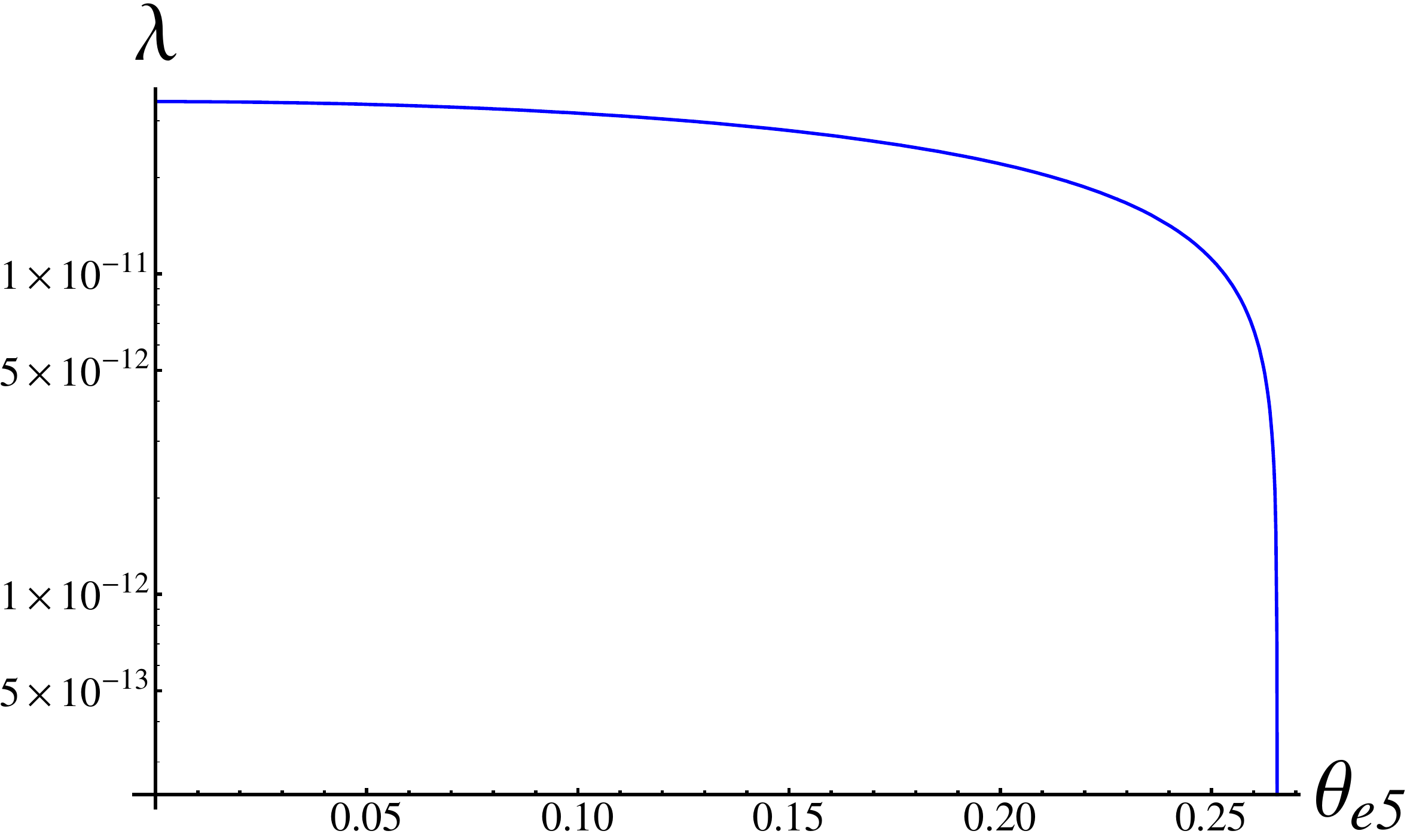}
	\caption{The constraint on the $\lambda$ coupling in the mixing angle $\theta_{e_5}$ (in radiant) space with fixed $m_5 = 10$~MeV, and above the curve region, there is no constraint on the $\lambda$ coupling.}
	\label{fig:lambdabbn-a}
\end{figure}

\begin{figure}
\centering
\includegraphics[scale=0.3, angle = 0]{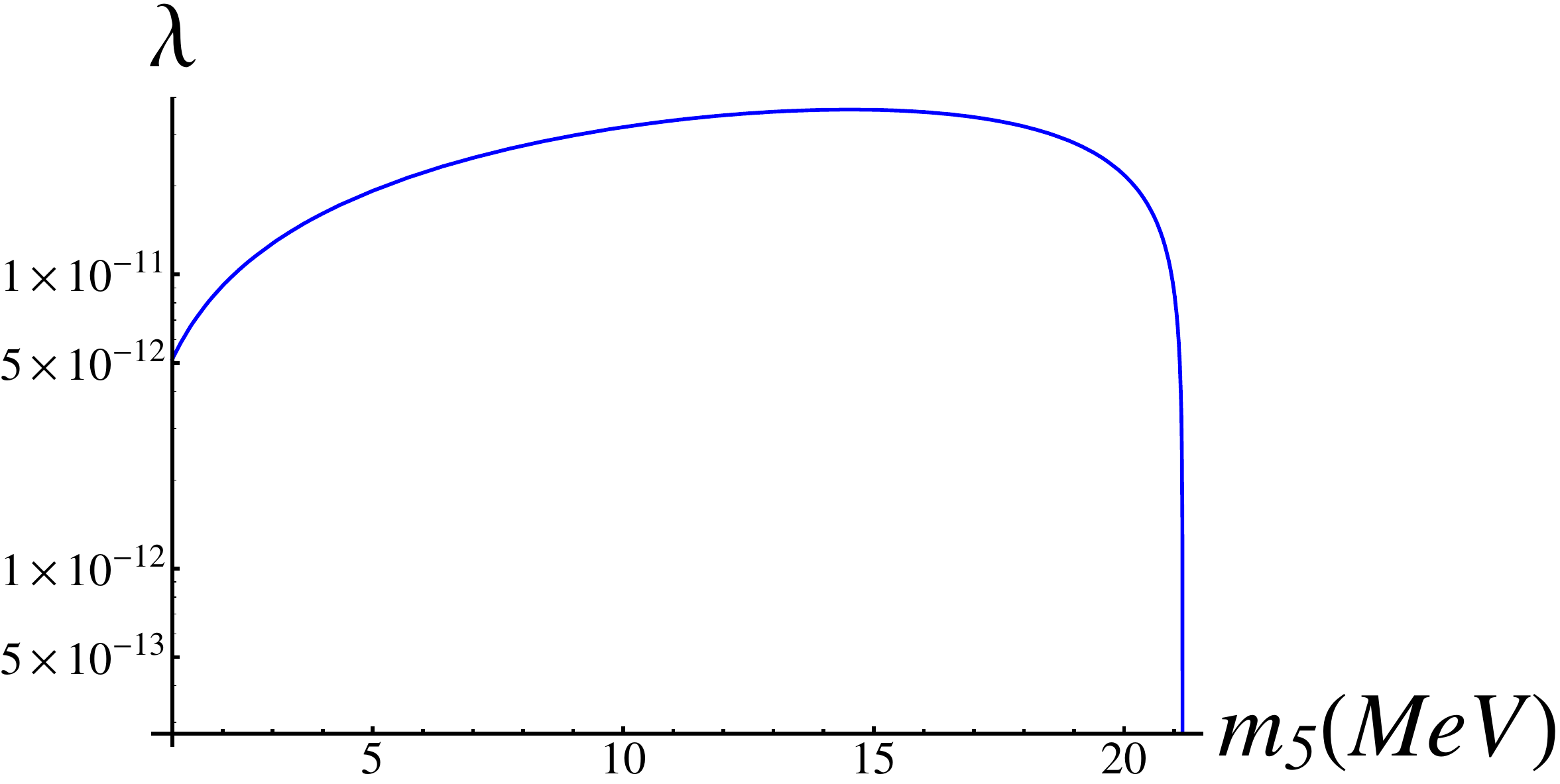}
	\caption{The constraint on the $\lambda$ coupling in the mass $m_5$ (in MeV) space, and above the curve region with fixed mixing angle $\sin\theta_{e5} = 0.1$, there is no constraint on the $\lambda$ coupling.}
	\label{fig:lambdabbn-b}
\end{figure}

In the mass region $m_5 \sim $[1, 34] MeV, the decay products of the $\nu_5 \rightarrow \nu_a + e^{+} + e^{-}$ can modify the Helium relic abundance. It puts stringent bounds on the lifetime of $\nu_5$, which is $\tau_5 > t_1 m_5^{\beta} + t_2$ where $t_1 = 1699, \; t_2 = 0.0544, \; \beta = -2.652$ for $\Delta N = 1$ and $t_1 = 1218, \; t_2 = 0.0513, \; \beta = -2.658$ for $\Delta N = 0.2$~\cite{Dolgov:2000jw, Dolgov:2003sg, Ruchayskiy:2012si}. The decay width of $\nu_5 \rightarrow \nu_a + e^{+} + e^{-}$ is,
\begin{equation}
\Gamma_{\nu_5}^{\mbox{\tiny{ee}}} \simeq \frac{\left(\frac{m_5}{10\text{MeV}}\right)^5 s_{2\theta_{e5}}^2 \hbar}{0.7} \;,
\end{equation} 
in which $s_{2\theta_{e5}} \equiv \sin2\theta_{e5}$. Taking the invisible decay into account, the constraint on the lifetime of $\nu_5$ further requires,
\begin{equation}
\frac{\hbar}{\Gamma_{\nu_5}^{\mbox{\tiny{inv}}} + \Gamma_{\nu_5}^{\mbox{\tiny{ee}}}} < t_1 \left(\frac{m_5}{\text{MeV}}\right)^{\beta} + t_2 \;,
\end{equation}
and, 
\begin{equation}
\frac{1}{16\pi \hbar} \lambda^2 m_5 > \frac{1}{t_1 \left(\frac{m_5}{\text{MeV}}\right)^{\beta} + t_2} - \frac{s_{2\theta_{e5}}^2}{0.7}\left(\frac{m_5}{10\text{MeV}}\right)^5 ,
\end{equation}
which means,
\begin{equation}
\lambda > 1.8\times 10^{-10} \sqrt{\frac{\frac{1}{t_1 \left(\frac{m_5}{\text{MeV}}\right)^{\beta} + t_2} - \frac{s_{2\theta_{e5}}^2}{0.7} \left(\frac{m_5}{10\text{MeV}}\right)^5}{\left(\frac{m_5}{\text{MeV}}\right)}} \;.
\end{equation}
The requirements on the $\lambda$ coupling can be further illustrated in the Fig.~\ref{fig:lambdabbn-a} and Fig.~\ref{fig:lambdabbn-b}. In Fig.~\ref{fig:lambdabbn-a}, the mass of $\nu_5$ is fixed to be $m_5 = 10$~MeV, and it shows the mixing angle $\theta_{e5}$ dependence while in Fig.~\ref{fig:lambdabbn-b}, the mixing angle $\theta_{e5}$ is fixed by $\sin\theta_{e5} = 0.1$ and it demonstrates the mass $m_5$ dependence.
Beyond the blue curve region, there is no constraint on the coupling $\lambda$ and note that the constraint on $\lambda$ from BBN is very weak.

\subsection{SN1987A bound}
The supernovae SN1987A observation of the neutrino energy emission puts strong bounds on the new particles interacting with active neutrinos~\cite{Raffelt:1987yt, Kainulainen:1990bn}. The region of $g_{\mbox{\tiny{min}}} < g < g_{\mbox{\tiny{max}}}$ (g is the coupling of the new interaction) has been ruled out and the lower bound $g_{\mbox{\tiny{min}}}$ origins from the total volume energy emission from the new particles while the $g_{\mbox{\tiny{max}}}$ comes from the blackbody surface emission rate. With no additional Yukawa interaction $\lambda \phi \nu_{4}\nu_{5}$, and the sterile neutrinos interact with the active neutrinos through mixings only,  for the lighter sterile neutrino $\nu_4$, the bound can be avoided due to the MSW resonance effects~\cite{Kainulainen:1990bn}. For the heavier sterile neutrino $\nu_5$ and the $\phi$ field, there are both the lower and higher bounds on the mixing angles with the active neutrinos. The dominant interaction of the sterile neutrino $\nu_5$ is the collision with nuclei through the mixing and the interaction rate is $\Gamma_{\mbox{\tiny{N}}\nu_5} = n_{\mbox{\tiny{N}}} \left<\sigma_{\mbox{\tiny{N}}\nu_5}\right> $ with $\left<\sigma_{\mbox{\tiny{N}}\nu_5}\right> = P(\nu_a \rightarrow \nu_5) \left<\sigma_{\mbox{\tiny{N}}\nu_a}\right>$ where $n_{\mbox{\tiny{N}}}$ is the nuclei number density in the supernovae, $\left<\sigma_{\mbox{\tiny{N}}\nu_5}\right>$ is the average collision cross section of $\nu_5$ and the nuclei while the $P(\nu_a \rightarrow \nu_5)$ is the oscillation probability between the active and sterile neutrinos, therefore,
\begin{equation}
\Gamma_{\mbox{\tiny{N}}\nu_5} \simeq \left(\frac{1}{2}\sin^22\theta_{m}\right) \times 7.5\frac{G_F^2 \left<E\right>^2}{\pi} \frac{\rho}{m_p} \;,
\label{eqn:gammanu5N}
\end{equation}
where $\theta_{m}$ is the mixing angle between the $\nu_5$ and $\nu_e$ in the nuclear matter, $G_F$ is the Fermi constant, $\left<E\right>$ is the average energy and we take $\left<E\right> = 100$~MeV and the supernovae core density $\rho = 8\times 10^{17}\text{kgm}^{-3}$ while $m_p$ is the proton mass.
Equivalently, the interaction length of $\nu_5$ in the supernovae is,
\begin{equation}
\ell_{\nu_5} \simeq 0.1m \left(\frac{100\text{MeV}}{\left<E\right>}\right)^2 \left(\frac{\rho_c}{\rho}\right) \frac{1}{\sin^22\theta_m}\;.
\label{eqn:lengthnu5N}
\end{equation}
in which $\rho_c = 2.6\times 10^{17}\text{kg m}^{-3}$ is the nuclear matter density.
If the mixing angles between the sterile neutrino and active neutrinos are small, the sterile neutrino can escape the supernovae core ($\sim 10$ km). The total energy loss rate over the whole volume requires that $\sin^2 2\theta_{m} \leq 7 \times 10^{-10}$~\cite{Kainulainen:1990bn}. On the other hand, if the $\nu_5$ is trapped inside the supernovae core, it radiates energy at its surface which is proportional to $T^4 R^2$ where T is the temperature and R is the radius of the $\nu_5$ surface inside the supernovae core and it thus is proportional to $\sqrt{\ell_{\nu_5}}$. If $\ell_{\nu_5}$ is larger than $\sim 1.5$m, it radiates too much energy and cools the supernovae too quickly and therefore it requires the mixing angle $\sin^2 2\theta_{m}\geq 2 \times 10^{-2}$~\cite{Kainulainen:1990bn}. The region of $7 \times 10^{-10} \leq \sin^2 2\theta_{m} \leq 2 \times 10^{-2}$ has been ruled out and similar but relaxed bounds exist for both $\nu_\mu$ and $\nu_\tau$ due to the fact that only neutral current interactions involved for $\nu_\mu$ and $\nu_\tau$ neutrinos. It means $1.5\text{m} \leq \ell_{\nu_5} \leq 4.6\times10^4$ km is ruled out, which can be generalized to other particles as well. Hence, no allowed parameter space to fit the LSND and MiniBooNe data. 

With the additional interaction $\lambda \phi \nu_{s1} \nu_{s2}$, the interaction lengths for the $\nu_4\;, \nu_5$ and $\phi$ can be short enough that they are trapped inside the supernovae core $\sim$1.5m by choosing proper $\lambda$ couplings. The new interaction rate of $\nu_4$ is,
\begin{equation}
\Gamma_{\nu_4} = \Gamma_{\mbox{\tiny{N}}\nu_4} + \Gamma_{\nu_4}^{\mbox{\tiny{dark}}} \; 
\label{eqn:gammanu4}
\end{equation}  
where the $\Gamma_{N\nu_4}$ is the interaction between the $\nu_4$ and background nuclei field and has the similar form as Eq.~(\ref{eqn:gammanu5N}) and the $\Gamma_{\nu_4}^{\mbox{\tiny{dark}}}$ are the interactions through the $2 \rightarrow 2$ scattering process between $\nu_4,\;\nu_5$ and $\phi$ in the dark sector. As for $\nu_5$, similarly to $\nu_4$, there are interactions between $\nu_5$ and the nuclei as well as the $2\rightarrow 2$ scattering in the dark sector. Additionally, $\nu_5$ can decay into $\nu_4,\; \phi$ and $\nu_a,\; e^+,\;e^-$, therefore, the interaction rate of $\nu_5$ is,
\begin{equation}
\Gamma_{\nu_5} = \Gamma_{\mbox{\tiny{N}}\nu_5} +\Gamma_{\nu_5}^{\mbox{\tiny{dark}}} +\Gamma_{\nu_5}^{\mbox{\tiny{ee}}} + \Gamma_{\nu_5}^{\mbox{\tiny{inv}}}\;.
\end{equation}

In terms of the $\phi$ field in the dark sector, the dominant interaction is through the scattering with $\nu_4\;,\nu_5$ in the dark sector.

In order to trap all of the light particles from the dark sectors to be within $\sim$ 1.5 m inside the supernovae core, we can require that the interaction length of $\phi$ field $\ell_{\phi} \lesssim 1.5$m which is the longest interaction length among $\nu_4,\;\nu_5$ and $\phi$. It leads to,
\begin{equation}
\Gamma_{\phi} \gtrsim \Gamma_{\mbox{\tiny{N}}\nu_5}\bigg|_{(\sin^22\theta_m = 2\times 10^{-2})}\;,
\end{equation}
where $\Gamma_{\phi} \simeq n_{\nu} \left(\frac{\lambda^4}{\left<E\right>^2}\right)$ and $\Gamma_{\mbox{\tiny{N}}\nu_5} \simeq n_{\mbox{\tiny{N}}} \frac{1}{2} \sin^22\theta_m G_F^2 \left<E\right>^2$. $n_{\nu}$ is the number density of the active neutrinos in the supernovae core and the light particles from the dark sector are in equilibrium with the active neutrinos. We know the number density ratio is roughly $n_{\nu}/n_{\mbox{\tiny{N}}} \simeq 0.05$, and it leads to,
\begin{equation}
n_{\nu} \frac{\lambda^4}{\left<E\right>^2} \gtrsim  n_{\mbox{\tiny{N}}} \frac{1}{2}\left(2\times10^{-2}\right) G_F^2 \left<E\right>^2 \;.
\end{equation}
After simplification, we arrive at,
\begin{equation}
\lambda \gtrsim \sqrt[4]{\frac{1}{0.05} \times 10^{-2} \times G_F^2}\left<E\right> \;,
\end{equation}
therefore,
\begin{equation}
\lambda \gtrsim 2\times 10^{-4}\;.
\end{equation}

On the other hand, the additional particles $\phi,\; \nu_4$ and $\nu_5$ scatter with the active neutrinos through the mixings between the sterile-active neutrinos and it modifies the mean free paths of the active neutrinos. If the interaction lengths of the active neutrinos are too short, the cooling time of the supernovae will be much longer than the standard scenario~\cite{Fayet:2006sa}. Therefore, we further require the scattering cross section between the new particle content in the dark sector and the active neutrinos is smaller than the scattering between the active neutrino and nucleon, which is on the order of weak interaction.
\begin{equation}
\left(\frac{1}{2}\sin^22\theta_m\right)^2 n_{\nu} \frac{\lambda^4}{\left<E\right>^2} \leq n_{\mbox{\tiny{N}}} G_F^2 \left<E\right>^2 \;,
\end{equation}
therefore, 
\begin{equation}
\lambda \leq \sqrt[4]{4\frac{n_{\mbox{\tiny{N}}}}{n_{\nu}} G_F^2} \frac{\left<E\right>}{\sin\theta_m} \;,
\end{equation}
and taking the mixing angle to be $\sin\theta_m \sim 0.1$, it requires $\lambda \lesssim 10^{-2}$. In summary, the allowed region of $\lambda$ coupling with the mixing angle $\sin\theta_m \sim \mathcal{O}(0.1)$ is,
 \begin{equation}
 2\times 10^{-4} \lesssim \lambda \lesssim 10^{-2}\;.
 \end{equation}

\section{dark matter}
\label{sec:three}
\subsection{Dark Matter Relic Abundance}
\begin{figure}[h]
\centering
\includegraphics[scale=0.9, angle = 0]{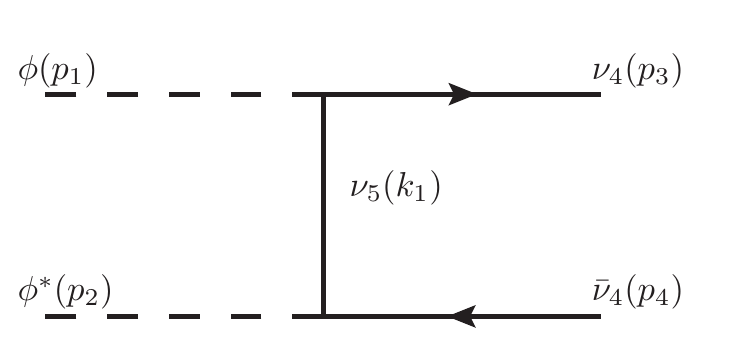}
	\caption{The $\phi$ field annihilation into the $\nu_4\bar{\nu}_4$ pairs through the t-channel $\nu_5$ exchange.}
	\label{fig:FeyDigDmAnnTUch}
\end{figure}

The dark matter $\phi$ field dominantly annihilates into the $\nu_4$ pairs through the $\nu_5$ exchange and it is shown in the Fig.~\ref{fig:FeyDigDmAnnTUch}. 
Using the non-relativistic limit and ignoring the mass of light sterile neutrino, $m_4$, 
the velocity averaged annihilation cross section through s-wave is,
\begin{equation}
\sigma_{\mbox{\tiny{ann}}} v_{\mbox{\tiny{rel}}} = \frac{\lambda^4}{4\pi} \frac{m_5^2}{(m_{\phi}^2 + m_5^2)^2} \simeq 0.3\left(\frac{\lambda}{10^{-3}}\right)^4\left(\frac{10\text{MeV}}{m_5}\right)^2\text{pb}\;.
\end{equation}
In order to generate the correct relic abundance, the relative velocity averaged annihilation cross section has to satisfy,
\begin{equation}
\sigma_{\mbox{\tiny{ann}}} v_{\mbox{\tiny{rel}}} \simeq 0.2 \times \frac{x_F}{\sqrt{g_{*}}} \left(\frac{\Omega_{dm} h^2}{0.11}\right)^{-1} \text{pb},
\end{equation}
in which $x_F = m_{\phi}/T_{F} \sim 12-19$ for particles in the MeV-GeV range and $g_{*} \sim \mathcal{O}(100)$ is the number of relativistic degrees of freedom. Hence, the allowed range of $\lambda$ with $m_5$ in [1, 34]~MeV is,
\begin{equation}
\lambda \in [3\times 10^{-4}, 2\times 10^{-3}]\;. 
\end{equation}
It is compatible with the BBN, supernovae SN1987A constraints.

\subsection{511 keV INTEGRAL Gamma Line}
\begin{figure}[t]
\includegraphics[scale=0.6, angle = 0]{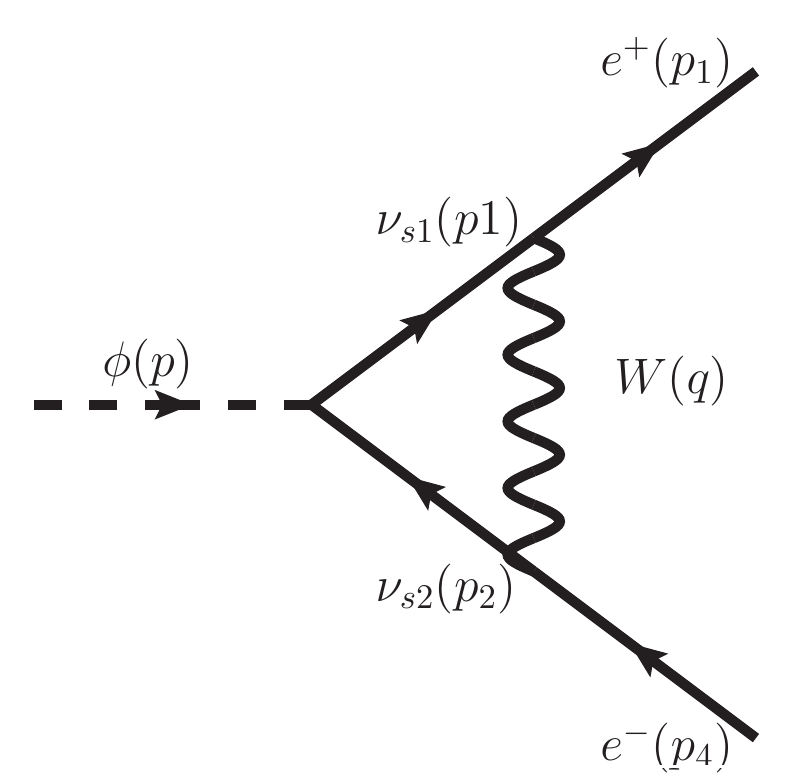}
	\caption{The $\phi$ field decays into the electron pairs through the triangle loop diagram.}
	\label{fig:loopdecayee}
\end{figure}

In addition, the $\phi$ field can decay into the light neutrino pairs at   tree level and at one loop, as well as the electron pairs through a loop if $m_{\phi} > 2m_{e}$ as   is shown in Fig.~\ref{fig:loopdecayee}.
Let us first calculate the decay width of the $\phi \rightarrow e^{+} e^{-}$. The amplitude of the triangle diagram in the Feynman-'t Hooft gauge can be written as,
\begin{eqnarray}
\mathcal{M} & \simeq & \int \frac{d^dq}{(2\pi)^d} \bar{u}^{s_3}(p_3) \frac{g_W^2}{2} (\lambda \sum_{i,j} U_{s1i} U_{ei}^* U_{s2j}^* U_{ej})  \\ \nonumber
& \times & \gamma^{\mu} P_L \frac{\Feyn{p}_1+m_i}{p_1^2-m_i^2}\frac{(\Feyn{p}_2 + m_j)}{p_2^2 - m_j^2} \gamma^{\nu} P_L\left[\frac{1}{q^2 - m_W^2}\right] \nu^{s_4}(p_4)\;,
\end{eqnarray}
where $P_L \equiv \frac{1-\gamma_5}{2}$ is the projection operator and $p_1 = p_3 + q,\; p_2 = p_4 - q,\; d = 4 - \epsilon\; (\epsilon \rightarrow 0)$. 
It is suppressed by the GIM cancellation mechanism, and the dominant contribution is from the term $m_i = m_j = m_5$ when $|U_{s15}|m_5 > m_4$, so we have,
\begin{eqnarray}
\mathcal{M}& \simeq & \left(\frac{g_W^2}{2}\right)\lambda U_{s15}U_{e5}^* U_{s25}^*U_{e5} m_5m_e\bar{u}^{s_3}(p_3) \\ \nonumber
& \times & \int \frac{d^dq}{(2\pi)^d}  \left[\frac{1}{(p_3+q)^2 (p_4-q)^2 (q^2 - m_W^2)}\right] \nu^{s_4}(p_4)\;,
\end{eqnarray}
Through the rotational transformation, we obtain,
\begin{eqnarray}
& & \int \frac{d^d q}{(2\pi)^d} \frac{1}{[(p_3 + q)^2][(p_4 - q)^2][q^2 - m_W^2]} \\ \nonumber
& =& 2 \int_0^1 dx_1 dx_2 \int \frac{d^d \ell}{(2\pi)^d} \frac{1}{[\ell^2 - \Delta]^3} \\ \nonumber
& = &\int_0^1 dx_1 dx_2 \frac{-i}{(4\pi)^2\Delta} \;,
\end{eqnarray}
with the following definitions,
\begin{eqnarray}
& & \ell  =  q + (x_1 p_3 - x_2 p_4) \;, \\ \nonumber
& & \Delta  =  m_W^2-(x_1+x_2)(m_W^2-m_5^2+m_e^2) \\ \nonumber
& & +(x_1^2+x_2^2)m_e^2-x_1x_2(m_{\phi}^2-2m_e^2)\;. 
\end{eqnarray}
Defining the integral $\int_0^1dx_1 dx_2 \frac{-i}{(4\pi)^2\Delta} \equiv I$, we have $I\simeq (2-3)\times10^{-3}\text{GeV}^{-4}$.
The decay width is approximately,
\begin{equation}
\Gamma_{\phi \rightarrow e^{+}e^{-}} \simeq 10^{-50} \left(\frac{\lambda}{10^{-3}}.\frac{|U_{s15}|}{10^{-7}}.\frac{m_5}{10\text{MeV}}\right)^{2}\left(\frac{m_{\phi}}{1\text{MeV}}\right) \;,
\end{equation}
where we have used $|U_{e5}^*U_{e5}|^2 \simeq 10^{-4}$ to be compatible with various sterile neutrino anomalies and $|U_{s25}|^2 \simeq 1$. In order to explain the 511 keV INTEGRAL gamma line, the lifetime of the $\phi$ dark matter needs to roughly follow $\tau_{\phi} \simeq 10^{18}$ years~\cite{Pospelov:2007xh}, under the assumption of the NFW dark matter density distribution, and more specifically, $\rho_{DM}(r) = \rho_0 \exp\left\{-\frac{2}{\alpha}\left[\left(\frac{r}{r_0}\right)^{\alpha}-1\right]\right\}$ with the parameter choice of $r_0 = 20h^{-1}$ kpc, $\rho_0$ being normalized by taking $\rho = 0.3\text{GeV}/\text{cm}^3$ at $r = 8.5$ kpc and $\alpha = $0.1 or 0.2~\cite{Navarro:2003ew, Picciotto:2004rp}, which means, $\Gamma_{\phi \rightarrow e^{+}e^{-}} \simeq 10^{-50}$ GeV and  
\begin{equation}
\label{eqn:lifetime}
 \left(\frac{\lambda}{10^{-3}}.\frac{|U_{s15}|}{10^{-7}}.\frac{m_5}{10\text{MeV}}\right)^{2}\left(\frac{m_{\phi}}{1\text{MeV}}\right) \sim \mathcal{O}(1) \;.
\end{equation}
Under the parameter choice of $m_5 \simeq 10$~MeV, $\lambda \simeq 10^{-3}$ and $m_{\phi} \simeq 1$~MeV, $|U_{s15}| \simeq 10^{-7}$, Eq.~(\ref{eqn:lifetime}) can be satisfied easily and $|U_{s15}|m_5 > m_4 (\simeq 0.5)\text{eV}$.

Aside from the $\phi \rightarrow e^{+}e^{-}$ through the triangle loop, the $\phi$ field can further decay into the light neutrino pairs ($\nu_i\bar{\nu}_i\;(i = 1,2,3,4)$) at the tree level through the active and strerile neutrino mixings. The decay rates are,
\begin{equation}
\Gamma_{\nu\nu}\simeq \frac{1}{16\pi}|U_{s1i}U_{s2i}^{*}|^2\lambda^2 m_{\phi}\; (i = 1, 2, 3, 4)\;,
\end{equation}
which need to be highly suppressed in order to have the $\phi$ field to be the dark matter candidate and explain the 511 keV INTEGRAL Gamma Line. More explicitly, $\Gamma_{\nu\nu} < 10^{-50}$ GeV, and it further puts an upper bound on the mixings. Now we can rewrite $\Gamma_{\nu\nu}$ as,
\begin{equation}
\Gamma_{\nu\nu}\simeq 10^{-50}\left(\frac{|U_{s1i}U_{s2i}^{*}|}{10^{-20}}\right)^2\left(\frac{\lambda}{10^{-3}}\right)^2 \left(\frac{m_{\phi}}{\text{MeV}}\right)\;,
\end{equation}
and  we can obtain $|U_{s1i}U_{s2i}^*| < 10^{-20}$ for $\lambda \simeq 10^{-3}$ and $m_{\phi} \sim \text{MeV}$. The $\phi$ field can also decay into the light neutrino pairs through a loop which is similarly highly suppressed and satisfies the constraints on the lifetime of the $\phi$ field.

\section{Mass Hierarchy and Mixing}
\label{sec:four}
The hierarchical mass structure of the sterile neutrinos and the mixing angles between active and sterile neutrinos can be generated through many mechanisms~\cite{Fan:2012ca, Langacker:1998ut, Sayre:2005yh, Berezhiani:1995yi,Dvali:1998qy,Merle:2013gea, Dinh:2006ia}. Generically, without tuning, mixing angles among Majorana neutrinos can be no larger than square roots of mass ratios, however special textures can allow mass hierarchies that are arbitrarily large with no constraints on mixing. Mixing between active neutrinos and heavy Dirac sterile neutrinos does not necessarily induce mass for the light neutrino states, so we will assume our  neutrinos are Dirac. We impose an additional $U(1)$ symmetry, which can generate the mass hierarchy and mixing angles among all of the SM fermions as well as the two sterile neutrinos. The $U(1)^{\prime}$ gauge symmetry is broken spontaneously at a high scale $\Lambda$, by the vacuum expectation value (VEV) of a flavon field $\eta$. All of the effective Yukawa matrices are generated through the higher dimensional operators,
\begin{equation}
\label{eqn:Yukawa1}
Y^{\mbox{\tiny eff}}_{ij} = \biggl( Y_{ij} \frac{\eta}{\Lambda} \biggr)^{|q_i + q_j + q_H|} = \biggl(Y_{ij} \epsilon \biggr)^{|q_i + q_j + q_H|}\, ,
\end{equation}
where $\eta$ is the flavon field and $q_i,\; q_j,\; q_H$ are the $U(1)^{\prime}$ charges of the $i$-th, $j$-th generation of the fermions as well as the Higgs respectively and $\epsilon = \frac{\left<\eta\right>}{\Lambda}$ with the choice of $\epsilon \simeq 0.22$, the Cabbibo angle. $\eta^{\prime}$ with opposite $U(1)^{\prime}$ charge of the $\eta$ field is inserted instead of $\eta$ field in the higher dimensional operator when $q_i+q_j+q_H < 0$. The $U(1)^{\prime}$ charges are normalized by $q_{\eta} = -1$. With proper charges for the fermions and scalars and $\mathcal{O}$(1) coefficients $Y_{ij}$ tuning, a realistic mass hierarchy and  weak mixing matrix is obtained, via the Froggatt-Nielson mechanism~\cite{Froggatt:1978nt}. Many examples can be found in~\cite{Chen:2006hn, Chen:2008tc, U1examples, Chen:2011sb} and we will follow closely to~\cite{Chen:2011sb}, in which realistic Dirac leptogenesis can also be realized naturally. The generation dependent broken $U(1)^{\prime}$ gauge symmetry leads to a massive gauge boson $Z^{\prime}$, and it is highly constrained by the flavor changing neutral current (FCNC) experiments and collider searches for the massive gauge boson $Z^{\prime}$~\cite{ATLAS:2013, CMS:2013}. Hence, the $U(1)^{\prime}$ breaking  is at a very high scale, which is also needed to generate the right amount of baryon number asymmetry (BAU). The $U(1)^{\prime}$ charges are further simplified by embedded $SU(5)$ symmetry and we follow the conventions in~\citep{Chen:2008tc}, therefore, the up-type quark masses can be generated by,
\begin{equation}
\epsilon^{|q_{t_i} + q_{t_j}+q_{H_1}|} {\bf 10}_i {\bf 10}_j {\bf 5}_{H_1} \;,
\end{equation}
where $q_{t_i},\; q_{t_j}$ denote the $U(1)^{\prime}$ charges of the $i$-th and $j$-th generations of the ${\bf 10}$ representation of the $SU(5)$ symmetry and $q_{H_1}$ is the $U(1)^{\prime}$ charge of the higgs field ${\bf 5}_{H_1}$ (we use $H_1$ in the following section). Similarly, the down-type quark masses are determined by,
\begin{equation}
\epsilon^{|q_{t_i}+q_{f_j}-q_{H_2}|} {\bf 10}_i {\bf \bar{5}}_j {\bf \bar{5}}_{H_2} \;,
\end{equation}
and $q_{f_j}$ is the $U(1)^{\prime}$ charge of the $j$-th generation of the ${\bf \bar{5}}$ representation of the $SU(5)$ symmetry and $q_{H_2}$ represents the $U(1)^{\prime}$ charge of the higgs field ${\bf 5}_{H_2}$ (we use $H_2$ in the following section). So the textures of Yukawa matrices of the up-type quark ($Y^u$), down-type quark ($Y^d$) and the charged lepton $Y^e = (Y^{d})^T$ are determined by the $U(1)^{\prime}$ charges $q_{t},\; q_{f},\; q_{H_1}$ and $q_{H_2}$. In the neutrino sector, it is slightly more complicated, all active and sterile neutrinos are Dirac fermions, the Majorana mass terms are forbidden by our $U(1)^{\prime}$ charge assignments. We take the 2 component Weyl spinor notation in this section, the three generations of the active neutrinos and its corresponding right-handed partners and the sterile neutrinos $\nu_{s_1},\nu_{s_2}$ in this letter can be expressed as,
\begin{eqnarray}
\nu_{d,i} \equiv \left(\begin{array}{c} \nu_{a, \alpha} \\ \nu_n^{\dagger \dot{\alpha}}\end{array}\right)_i, \; \nu_{s,j} \equiv \left(\begin{array}{c} \nu_{l, \alpha} \\ \nu_r^{\dagger \dot{\alpha}}\end{array}\right)_j\;\left(\begin{array}{c}i = 1, 2, 3\\ j = 1, 2\end{array}\right)\;.
\end{eqnarray}
We choose the  lepton number of $\nu_{d,i}$ and $\nu_{s,j}$ to be$+1$. The neutrino masses are generated by coupling to scalar fields with vevs, as described in the Lagrangian terms:
\begin{eqnarray}
M_{\nu}  & = Y^{an}\frac{<\Xi >}{\Lambda} \nu_a\nu_n \left<H_1\right> + Y^{ar}\nu_a\nu_r\left<H_1\right> \\ \nonumber
&+ Y^{ln}\nu_{l}\nu_n \left<\chi\right> + Y^{lr} \nu_l \nu_r\left<\chi\right> + h.c.\;. 
\end{eqnarray}
The flavon field $\Xi$ acquires a VEV upon the $U(1)^{\prime}$ symmetry breaking and $\frac{\langle \Xi \rangle}{\Lambda} \sim \mathcal{O}(1)$. The flavon fields $\Xi$ and $\xi$ with opposite $U(1)^{\prime}$ charge of $\Xi$ play an important role in the Dirac Leptogenesis mechanism, as it is shown in~\cite{Chen:2011sb}. The neutrino Yukawa matrices are,  
\begin{eqnarray}
Y_{ij}^{an} & \simeq & \epsilon^{|q_{f_i} + q_{n_j} + q_{H_1} + q_\Xi|} \, , \\ \nonumber
Y_{ij}^{ar} & \simeq & \epsilon^{|q_{f_i} + q_{r_j} + q_{H_1}|}  \, , \\ \nonumber
Y_{ij}^{ln} & \simeq & \epsilon^{|q_{l_i} + q_{n_j} + q_{\chi}|} \, , \\ \nonumber
Y_{ij}^{lr} & \simeq & \epsilon^{|q_{l_i} + q_{r_j} + q_{\chi}|} \, .
\end{eqnarray}
in which $q_{n_j},\; q_{l_i}$ and $q_{r_j}$ are the $U(1)^{\prime}$ charges of the $j$-th generation of $\nu_n$, $i$-th generation of $\nu_l$ and $j$-th generation of $\nu_r$ respectively and $q_{\Xi},\;q_{\chi}$ are the $U(1)^{\prime}$ charges of the flavon field, $\Xi$ and $\chi$. The $\chi$ field gets a VEV $v_{\chi} \sim \mathcal{O}(\text{MeV})$ as the $U(1)^{\prime}$ symmetry is broken.
Consequently, the mass matrix of all neutrinos can be written as,
\begin{eqnarray}
M^{\nu} = \left(\begin{array}{cc} Y^{an}\frac{<\Xi>}{\Lambda}\left<H_1\right>  \;\; & Y^{ar}\left<H_1\right> \\ Y^{ln}\left<\chi\right> & Y^{lr}\left<\chi\right>\end{array}  \right) +h.c. \; .
\end{eqnarray}
Among the active neutrinos, the mixings are large and it roughly follows the nearly Tri-bimaximal mixing with non-zero $\theta_{13}$. However, the mixings between active and sterile neutrinos are small.     In order to fit   anomalous neutrino experimental results such as LSND, MiniBooNE, the mixings $|U_{e_i}|\;(i=4,5)$ and $|U_{\mu_i}| \;(i=4,5)$ need to be on the order of $\mathcal{O}(0.1)$.  In addition, the decay channel of $\phi\rightarrow \nu_i\bar{\nu}_i$ requires $|U_{s1i}U_{s2i}^*|<10^{-20}\;(i=1,\;2,\;3,\;4)$, while $|U_{s15}|\sim 10^{-7}-10^{-5}$ is needed to explain the 511 keV INTEGRAL results. So the mixings between active and sterile neutrinos as well as between two sterile neutrinos (as defined by the difference between the U(1) charge basis and the mass eigenstate basis) are generated by another mechanism, such as radiative corrections. Therefore, the Yukawa matrices $Y^{ar}$ and $Y^{ln}$ as well as the $Y_{12}^{lr}$ and $Y_{21}^{lr}$ terms are forbidden with our designated $U(1)^{\prime}$ charges due to the fact that only the terms with the number of the inserted $\eta$ field being integer are allowed.

Furthermore, we choose $q_{\phi} = 0$ so that the $\phi$ field is a gauge singlet under the $U(1)^{\prime}$ symmetry and the $\lambda$ coupling can be generated by the higher dimensional operator,
\begin{equation}
\lambda \simeq \epsilon^{|q_{l_1}+q_{r_2}|} \simeq \epsilon^{|q_{l_2}+q_{r_1}|}\;.
\end{equation}

There are 17 free parameters in total by far. It can be further reduced by anomaly cancellation conditions. Given the simplified $U(1)^{\prime}$ charge assumption inspired by $SU(5)$ GUT symmetry, the six anomaly cancellation constraints relating to the $U(1)^{\prime}$symmetry are reduced down to the following three,
\begin{eqnarray}
[SU(5)]^{2}U(1)^{\prime}:  \sum_{i=i}^{3} (\frac{1}{2} q_{f_i} + \frac{3}{2} q_{t_i}) = 0 \; , \\ 
\mbox{gravity}-U(1)^{\prime}:  \sum_{i=1}^{3} (5 q_{f_i} + 10 q_{t_i} + q_{n_i}) \\ \nonumber
+ \sum_{i=1}^{2} (q_{l_i} + q_{r_i}) = 0 \; , \\
U(1)^{\prime 3}:  \sum_{i=1}^{3} (5q_{f_i}^3 + 10 q_{t_i}^3 + q_{n_i}^3) \\ \nonumber
 + \sum_{i=1}^{2} (q_{l_i}^3 + q_{r_i}^3) + q_{\text{exotic}}^3 = 0 \; .
\end{eqnarray}
In this model, the mixed anomalies $[SU(5)]^{2}U(1)^{\prime}$ and $\mbox{gravity}-U(1)^{\prime}$ are cancelled among the SM fermions and sterile neutrinos and the $U(1)^{\prime 3}$ anomaly is cancelled with the additional heavy exotic fermions, which are SM singlets. Inspired by the charge splitting parametrization~\cite{Chen:2006hn}, we parametrize the $U(1)^{\prime}$ charge as following,
\begin{eqnarray} 
\begin{array}{lll} q_{t_1} & = & -\frac{1}{3} q_{f_1} - 2a \; , \nonumber \\
q_{t_2} & = & -\frac{1}{3} q_{f_2} + a + a^{\prime} \; ,  \nonumber \\
q_{t_3} & = & -\frac{1}{3} q_{f_3} + a - a^{\prime} \; ,  \end{array} \nonumber \quad 
\begin{array}{lll} q_{n_1} & = & x -\frac{5}{3} q_{f_1} - 2b \; , \nonumber  \\
q_{n_2} & = & x -\frac{5}{3} q_{f_2} + b + b^{\prime} \; , \nonumber \\
q_{n_3} & = & x -\frac{5}{3} q_{f_3} + b - b^{\prime} \; , \end{array}
\label{eqn:param}
\end{eqnarray}
and require $q_{l_1} + q_{l_2} + q_{r_1} + q_{r_2} = -3x$ so that the $[SU(5)]^2U(1)^{\prime}$ and gravity-$U(1)^{\prime}$ conditions are automatically satisfied.

Similarly to~\cite{Chen:2011sb}, we impose various relations among the $U(1)^{\prime}$ charges which are well-motivated by the observed hierarchy and mixing patterns: 

$\bullet$ little or no suppression in the corresponding mass terms since the third generation of quarks, charged lepton are heavy, we assume,
\begin{equation}
|2q_{t_3} + q_{H_1}| = 0 \, , \quad |q_{t_3} + q_{f_3} - q_{H_2}| = 2 \, .
\end{equation}

$\bullet$ in order to obtain nearly tri-bimaximal mixing pattern in the neutrino sector, we also require,
\begin{eqnarray}
q_{f_2} & = & q_{f_3} \, , \, |q_{f_1} - q_{f_2}| = 1 \, , \nonumber\\ 
q_{n_2} & = & q_{n_3} \, , \, |q_{n_1} - q_{n_2}| = 1 \,, \, (b = -8/9 \,, b^{\prime} = 0).
\end{eqnarray}

$\bullet$ we choose $a = -7/9$ and $a^{\prime} = 1$ for the realistic quark and charged lepton masses and mixings to be consistent with experimental data.

$\bullet$ In order to generate the correct mass hierarchy for neutrinos, we further require the suppression powers to satisfy,
\begin{eqnarray}
p_{22} = 20,\; p_{44} = 8,\; p_{55} = 0,
\end{eqnarray}
in which $Y^{\nu}_{ij} \simeq (\epsilon)^{p_{ij}}$. It can be converted into the requirements on the $U(1)^{\prime}$ charges, which are,
\begin{eqnarray}
&\frac{8}{3}+x+q_{\Xi} = 20,\; q_{l_1}+q_{r_1}+q_{\chi} = 8,\; \\ \nonumber 
&3x-q_{l_1}-q_{r_1}+q_{\chi} = 0.
\end{eqnarray}

$\bullet$ The coupling $\lambda$ is also generated by the higher dimensional operator, therefore, we require,
\begin{equation}
q_{l_2}+q_{r_1} = 5,\; q_{l_2} = q_{l_1}+\frac{1}{2}.
\end{equation}

$\bullet$ By far, there is only one free parameter $q_{l_1}$ and we further choose $q_{l_1} = 9/2$ to obtain a relatively small cubic anomaly from all SM fermions and additional neutrinos, which means, $ \sum_{i=1}^{3} (5q_{f_i}^3 + 10 q_{t_i}^3 + q_{n_i}^3) + \sum_{i=1}^{2} (q_{l_i}^3 + q_{r_i}^3)= -\frac{13}{9}$ and it can be further cancelled by the additional heavy exotic fields. 
\begin{table}[t!]
\begin{tabular}{l|l||l|l}\hline\hline\
Field & $U(1)^{\prime}$ charge & Field & $U(1)^{\prime}$ charge \\ \hline
${\bf \bar{5}}_1$& $q_{f_1} = -13/3$ & ${\bf 10}_1$ & $q_{t_1} = 3$ \\ \hline
${\bf \bar{5}}_2$& $q_{f_2} = -16/3$ & ${\bf 10}_2$ & $q_{t_2} = 2$ \\ \hline
${\bf \bar{5}}_3$& $q_{f_3} = -16/3$ & ${\bf 10}_3$ & $q_{t_3} = 0$\\ \hline   
$\nu_{n_1}$& $q_{n_1} = 28/3$ & $\nu_{r_1}$ & $q_{r_1} = 0$ \\ \hline
$\nu_{n_2}$& $q_{n_2} = 25/3$ & $\nu_{r_2}$ & $q_{r_2} = -17/2$ \\ \hline   
$\nu_{n_3}$& $q_{n_3} = 25/3$ & $\nu_{l_1}$ & $q_{l_1} = 9/2$ \\ \hline  
$\eta$ & $q_{\eta} = -1$ & $\nu_{l_2}$ & $q_{l_2} = 5$ \\ \hline
$H_1$ & $q_{H_1} = 0$ & $\Xi$ & $q_{\Xi} = 17$ \\ \hline
$H_2$ & $q_{H_2} = -22/3$ & $\chi$ & $q_{\chi} = 7/2$ \\ \hline \hline
\end{tabular}
\caption{The $U(1)^{\prime}$ charges of the particles in the model.}   
\label{tbl:u1Charge2}
\end{table}
With the $U(1)^{\prime}$ charge assignments summarized in Table~\ref{tbl:u1Charge2}, it leads to the following effective mass matrix for the up-type quarks,
\begin{eqnarray}
\label{eqn:upYukawaNew}
M^u & \sim & \left(\begin{array}{ccc} \epsilon^{6} & \epsilon^{5} & \epsilon^{3} \\ \epsilon^{5} & \epsilon^{4} & \epsilon^{2} \\ \epsilon^{3} & \epsilon^{2} & \epsilon^{0} \end{array} \right) \left<H_1\right> \; ,
\end{eqnarray}
and the effective mass matrices of the down-type quarks and thus charged leptons are given by,
\begin{eqnarray}
\label{eqn:downYukawaNew}
M^d & \simeq & \left(\begin{array}{ccc} \epsilon^{6} & \epsilon^{5} & \epsilon^{5} \\ \epsilon^{5} & \epsilon^{4} & \epsilon^{4} \\ \epsilon^{3} & \epsilon^{2} & \epsilon^{2} \end{array} \right) \left<H_2\right> \; ,
\end{eqnarray} and
\begin{equation}
M^e \simeq (M^{d})^{T} \; .
\end{equation}
The effective neutrino Dirac mass matrix is given by, 
\begin{eqnarray}
M^{an} & \simeq & \left(\begin{array}{ccc}\epsilon^{22} & \epsilon^{21} & \epsilon^{21} \\ \epsilon^{21} & \epsilon^{20} & \epsilon^{20} \\ \epsilon^{21} & \epsilon^{20} & \epsilon^{20} \end{array}\right) \left<H_1\right> \;, \\ \nonumber
M^{lr} & \simeq & \left(\begin{array}{cc}\epsilon^{8} & 0 \\ 0 & 1 \end{array}\right) \left<\chi\right> \;.
\end{eqnarray}
After the proper $\mathcal{O}$(1) coefficients tuning, all fermion masses and mixing angles can be accommodated. 

\section{Conclusion}
\label{sec:five} 
We have shown in this letter that with the additional interaction $\lambda \nu_4 \nu_5 \phi$, we can avoid various collider and cosmological bounds such as supernovae SN1987A and BBN observation with a better fit to the LSND and MiniBooNe results. The additional invisible decay $\nu_5 \rightarrow \nu_4 \phi$ can eliminate the BBN bound, which is weak compared to supernovae bound. The light particles in the dark sector scatter strongly among themselves so that they are all trapped in the inner core of supernovae so as to avoid the supernovae SN1987A bound. With the same additional interaction, the $\phi$ field can naturally be our MeV dark matter candidate and explain the 511 keV INTEGRAL gamma line. Different from the mechanism that $\phi$ field annihilates into SM fermion pairs through the light vector boson (U boson), the $\phi$ field dominantly annihilates into the sterile neutrino $\nu_4\bar{\nu}_4$ pair through the exchange of the $\nu_5$. It gives rise to the correct relic abundance while explain the 511 keV INTEGRAL gamma line naturally through the dark matter decaying into the electron pairs at the loop level. In addition, all fermions masses and mixings can be naturally generated through the $U(1)^{\prime}$ family symmetry with a bonus of the realistic Dirac Leptogenesis to explain the BAU.  
\begin{acknowledgments}
JH is supported by the DOE Office of Science and the LANL LDRD program. JH thanks Zhongbo Kang and Haibo Yu for   helpful disscussions and the hospitality of the University of Washington in Seattle where part of the work was done. AN acknowledges partial support from the Department of Energy under grant number DE-FG02-96ER40956.
\end{acknowledgments}

\end{document}